# Grain boundary complexions and the strength of nanocrystalline metals: Dislocation emission and propagation

Vladyslav Turlo[a], Timothy J. Rupert[a,b]*

[a] Department of Mechanical and Aerospace Engineering, University of California, Irvine, CA 92697, USA

[b] Department of Chemical Engineering and Materials Science, University of California, Irvine, CA 92697, USA

**Abstract**

Grain boundary complexions have been observed to affect the mechanical behavior of nanocrystalline metals, improving both strength and ductility. While an explanation for the improved ductility exists, the observed effect on strength remains unexplained. In this work, we use atomistic simulations to explore the influence of ordered and disordered complexions on two deformation mechanisms which are essential for nanocrystalline plasticity, namely dislocation emission and propagation. Both ordered and disordered grain boundary complexions in Cu-Zr are characterized by excess free volume and promote dislocation emission by reducing the critical emission stress. Alternatively, these complexions are characterized by strong dislocation pinning regions that increase the flow stress required for dislocation propagation. Such pinning regions are caused by ledges and solute atoms at the grain-complexion interfaces and may be dependent on the complexion state as well as the atomic size mismatch between the matrix and solute elements. The trends observed in our simulations of dislocation propagation align with the available experimental data, suggesting that dislocation propagation is the rate-limiting mechanism behind plasticity in nanocrystalline Cu-Zr alloys.

## 1. Introduction

Nanocrystalline metals demonstrate improved mechanical properties such as high strength [1], fatigue resistance [2] and wear resistance [3], but are usually limited by their low ductility. For example, Wang et al. [4] performed tensile experiments of nanocrystalline Cu with a grain size of ~30 nm, made by surface mechanical attrition treatment. These authors found that the yield strength of this material was 760 MPa, a very high value for Cu, while the plastic strain-to-failure was only 3%, an extremely low value. Similar values of yield stress and strain-to-failure of 740 MPa and 4%, respectively, were obtained by Khalajhedayati et al. [5] for nanocrystalline Cu with a grain size of 30 nm made by ball milling. However, Khalajhedayati et al. [5] were able to improve the ductility of nanocrystalline Cu-Zr alloys by introducing disordered grain boundary complexions [6-8], with strain-to-failure values of 56% observed, representing an order of magnitude improvement. These authors created nanocrystalline samples with three complexion types: (1) clean grain boundaries (CGBs), (2) ordered grain boundary complexions (OGBCs), and (3) amorphous intergranular films (AIFs). CGBs have no alloying elements at the grain boundaries and, in this case, the sample was pure nanocrystalline Cu. In contrast, OGBCs and AIFs are formed because of solute segregation to grain boundaries, being fabricated in Cu-Zr in the work of Khalajhedayati et al. OGBCs demonstrate the same structural order as the original grain boundaries but have segregating dopants, while AIFs exist as thin amorphous films with thicknesses on the order of a few nanometers. Pan and Rupert [9, 10] provided a mechanistic explanation for the improved ductility by studying the effect of CGBs and AIFs on dislocation absorption at grain boundary sites, finding a greater ability of AIFs to absorb dislocations before crack nucleation. Moreover, this absorption ability was directly related to the thickness of the AIFs, with thicker AIFs enabling more ductility.

Khalajhedayati et al. also saw that the introduction of OGBCs and AIFs leads to increased strength, with Cu-Zr alloys with AIFs having yield strengths in excess of 1 GPa. However, the mechanism behind this improved strength has not been described. In general, a material's strength is controlled by the dominant deformation mechanism behind plastic flow. In nanocrystalline materials with grain sizes of several tens of nanometers, the traditional metallic deformation mechanism of easy dislocation slip through the grain interior becomes suppressed by the nanometer-scale grain size and plastic deformation begins to be controlled by grain boundary-dislocation interactions [11]. As illustrated in Figure 1, dislocations are nucleated from [12, 13], propagate along [11] and subsequently absorbed by [14] grain boundaries. While dislocation absorption at grain boundaries impacts the nanocrystalline material's ductility [9], the dislocation emission and dislocation propagation processes are both candidates for controlling a nanocrystalline material's strength [11]. However, since both are observed during plastic deformation of nanocrystalline metals and alloys [15, 16], it remains unclear which is the rate-limiting deformation mechanism controlling plasticity in these materials.

In fact, two different types of phenomenological models, each based on one of these mechanisms, have been developed to try to model the yield strength of nanocrystalline materials. For example, Asaro et al. [17] proposed a model based on dislocation emission from the grain boundaries as the main deformation mechanism. Due to the nanometer-scale grain size, Asaro et al. assumed that the leading partial dislocation is emitted from the grain boundary interface, propagates through the grain interior, and then reaches the other side of the grain before the trailing partial dislocation is emitted. According to this model, the nanocrystalline material's strength is directly proportional to shear modulus and stacking-fault energy. In turn, both of these properties would depend on solute concentration, linking a nanocrystalline alloy's strength with the composition of the grain interior. Tucker and McDowell [18] studied how nonequilibrium grain boundary structure affects dislocation nucleation using molecular dynamics (MD) simulations, finding that excess free volume reduces the critical stress for dislocation nucleation and emission. Borovikov et al. [19] investigated the effect of solutes on dislocation nucleation from $\Sigma 11$ grain boundaries using Monte Carlo (MC) and MD simulations, finding that the addition of Ag to Cu and Cu to Ag may increase the yield stress associated with this deformation mechanism. In contrast, other phenomenological models have been developed that focus on dislocation propagation. For example, Yamakov et al. [16] studied the effect of stacking fault energy on plastic deformation in nanocrystalline metals using MD simulations, finding that, for relatively high stacking fault energies, the strength-limiting mechanism is the emission of a pair of Shockley dislocations that traverses a grain. Van Swygenhoven et al. [11] demonstrated that the partial dislocation propagation in nanocrystalline metals can be slowed down or even stopped by bowing between grain boundary pinning sites such as grain boundary ledges. To show how these models would be affected by the addition of alloying elements, Rupert et al. [20] developed a model for solid solution strengthening that incorporated dislocation propagation as the critical step. This model was then used to describe a wide range of material behaviors in the literature, ranging from strong strengthening to solid solution softening. While both types of strength models can provide insight into the effects of doping, neither have treated the possibility of disordered complexions and their effect on strength to date.

In this paper, we use MD and hybrid molecular dynamics/Monte Carlo (MD/MC) methods to perform atomistic simulations of dislocation emission and propagation in samples with a variety of grain boundary complexion states. The aim of our work is to (1) determine how complexions affect each mechanism and (2) isolate the dominant deformation mechanism that controls plasticity in nanocrystalline metals with grain sizes of several tens of nanometers. To do so, we consider the same bicrystal sample geometry for both dislocation emission and propagation simulations. This allows us to isolate and study these deformation mechanisms one-by-one, minimizing possible sources of error. We find that OGBCs and AIFs promote dislocation emission by reducing the critical emission stress due to an excess of free volume caused by Zr segregation. Alternatively, OGBCs and AIFs both increase the average stress required for dislocation propagation, thanks to large ledges in boundary structure as well as chemical effects, both of which make the dislocation pinning sites stronger. Finally, we compare our simulation results with experimental results available in the literature, finding that the effect on dislocation propagation is consistent with observed strengthening trends. Consequently, we conclude that dislocation propagation is the limiting deformation mechanism that controls plasticity in nanocrystalline metals with grain boundary complexions.

## 2. Computational Methods

MD and hybrid MD/MC simulations were performed with the Large-scale Atomic/Molecular Massively Parallel Simulator (LAMMPS) software [21] using an embedded-atom method (EAM) interatomic potential for the Cu-Zr system [22]. Cu and Cu-Zr samples were chosen to match existing materials available in the experimental literature. Analysis and visualization of the simulation results was performed with the OVITO tool [23]. Dislocations and their Burgers vectors were identified using the Dislocation Extraction Algorithm (DXA) [24]. The local atomic structure was determined by adaptive common neighbor analysis, as implemented in OVITO, with FCC atoms appearing green, HCP atoms appearing red, BCC atoms appearing blue and all other atoms appearing gray. To quantify the boundary roughness, we define the roughness parameter $r = (S_i - S_{YZ})/S_{YZ}$, where $S_i$ is the average surface area of the interface and $S_{YZ}$ is the surface area of a flat interface. A surface mesh for grain boundary interfaces was constructed by using the corresponding algorithm implemented in OVITO *[25]*, with a probe sphere radius of 0.6 nm and zero smoothing level. To study dislocation emission and propagation, a bicrystal simulation cell with the two 90° twist grain boundaries was considered (Figure 2). The [111] slip plane and the (110) direction are the most preferable slip system for an edge dislocation in an fcc crystal, so the inner grain was set up as shown in Figure 2(a). The slip plane is perpendicular to the Z-direction and an edge dislocation will glide along the Y-direction. The crystal orientation of the outer grain was chosen to create high-angle grain boundaries, because experimental observations demonstrate that such boundaries hamper direct dislocation transmission across the grain boundary [26]. A defect-free bicrystal, shown in Figure 2(a), was used to study dislocation emission, while two edge dislocations of opposite character were inserted into the inner grain to study dislocation propagation, as shown in Figure 2(b). Two pairs of partial dislocations with stacking faults between each pair were first created at the center of the inner grain in Y-direction by removing one-half of the XZ atomic plane, followed by relaxation of the system with a molecular statics method at 0 K. After this relaxation, the inner grain was joined to an outer grain. The choice of the asymmetric high-angle grain boundaries made it necessary to adjust the intergranular separation distance and to perform the small shifts of the one grain in relation to another in Y- and Z-directions to minimize the grain boundary energy. Multiple initial configurations were created as proposed by Tschopp et al. [27] and molecular statics at a constant zero pressure was applied to obtain a final state for each configuration. The minimum energy configuration was chosen for further simulation tasks.

The simulation cells were ~36 nm (two grains of 18 nm width) in the X direction, ~31 nm in the Y direction, and ~14 nm in the Z direction. The cell length in the X direction was chosen to be large enough to minimize the effect of periodic boundary conditions on dislocation nucleation and emission [28]. Following the same logic, a relatively large cell size was considered in the Y-direction to allows us to determine the dislocation propagation velocity with more accuracy. The length of the simulation box in Z direction was chosen to be 2-2.5 times smaller than in other directions for computational efficiency.

Simulation cells with and without the manually inserted dislocations were set up at 0 K with the corresponding lattice parameter for Cu. Periodic boundary conditions were applied in all directions, while a Nosé-Hoover thermostat and a Parrinello-Rahman barostat were used for all simulations. The thermostatting and barostatting times were adjusted to be 0.1 and 1.0 ps, respectively, and the integration time step was chosen to be 1 fs. The simulation cells were first heated to 100 K with a temperature ramp of 10 K/ps and then equilibrated at this temperature for additional 100 ps in an isothermal-isobaric (NPT) ensemble under zero pressure. This procedure allowed us to obtain the samples with two CGBs.

To obtain the samples with OGBCs and AIFs, the CGB samples were first heated up to an equilibration temperature, then the equilibrium boundary states were determined with the hybrid MD/MC method at a given composition, and finally the equilibrated samples were cooled down to 100 K with a temperature ramp of 10 K/ps. A variance-constrained semi-grand canonical ensemble [29] was used to perform an MC step, followed by relaxation in an NPT ensemble for 0.1 ps under zero pressure at a given temperature. This procedure was repeated until the equilibrium configuration was reached. The simulation was stopped when the absolute value of the slope of the potential energy over the last 400 ps of MD simulation was less than 1 eV/ps, following the criteria used by Pan and Rupert [30] in their work on grain boundary complexion formation in Cu-Zr.  After equilibration and cooling, each system was relaxed at 100 K for an additional 150 ps.  For statistical purposes, six thermodynamically equivalent configurations of each system were stored (every 10 ps) during the last 50 ps of the equilibration process.

To obtain complexion states that mimic the variety observed in experimental studies, samples with global compositions in a range from 0.1 to 2 at.% Zr were considered in the hybrid MD/MC simulations.  Two equilibration temperatures of 300 K and 1200 K were studied. The temperature of 300 K was chosen to be low enough to obtain OGBCs in a doped Cu-Zr sample.  A composition of 0.3 at.% Zr was determined as the highest composition that would allow for the creation of fully doped interfaces while retaining the structural order found in the original, clean grain boundary.  The OGBC samples are equivalent to the experimentally observed ordered grain boundary interfaces that have been reported in some nanocrystalline Cu-Zr systems [5, 31, 32].  These materials were annealed at an intermediate or high temperature to allow for Zr segregation to the grain boundaries, then slowly cooled down to room temperature.  An additional sample with a global composition of 1.3 at.% Zr and equilibrated at 1200 K was created to allow for the formation of AIFs.  In the end, this specimen contained two AIFs that were each ~2 nm thick.  Representative interfaces for the three samples are shown in Figure 3.  The OGBC has a structure that is similar to the CGB but with added Zr dopants, whereas the AIF has a disordered structure with rough interfaces between the amorphous nanolayer and the surrounding grains.  The distribution of the solute atoms in the grain interior can depend on the global composition and annealing temperature. The OGBC samples, which were created at the low temperature of 300 K, demonstrate complete segregation of the Zr atoms to grain boundaries.  In contrast, the AIF samples experience dsegregation to AIFs along with some random distribution of Zr atoms inside the grains, which can be attributed to the increase in Zr solubility in Cu as temperature increases. However, solute atoms may influence the dislocation propagation process by altering the Young's modulus or equilibrium lattice constant of the grain interior [20] or even reducing stacking fault energy [26]. To avoid such effects and keep our focus on changes to grain boundary structure alone, the Zr atoms inside the inner grain were removed from each sample containing pre-inserted dislocations.

To study dislocation emission and propagation processes, tensile and shear deformations of the simulation cell were performed using the non-equilibrium molecular dynamics (NEMD) method [33, 34]. A $10^9$ s$^{-1}$ engineering strain rate was applied in all of the simulations for computational efficiency and to allow for multiple simulations of each situation, improving statistics. The tensile deformation to probe dislocation emission was carried out in a quasi-static manner [28], following two steps: (1) tension of the sample along the X direction over 1 ps and (2) relaxation of the sample in NPT ensemble under zero pressure along the Y and Z directions over 2 ps. The two steps above were repeated until 10% tensile strain was reached. The quasi-static approach allows for the identification of the exact time of dislocation emission, is computationally efficient, and has only a very small dependence on a tensile strain rate. Spearot et al. [35] used such a method to study dislocation nucleation in Cu single crystals by deforming the system along several crystallographic orientations and found only a 4% variation in the critical nucleation stress when varying strain rate from $10^7$-$10^9$ s$^{-1}$. A YZ shear deformation of the simulation cell with pre-inserted dislocations was continuously performed at 100 K over 100 ps to reach a 10% shear strain during the dislocation propagation simulation. Due to the short time scales associated with the MD method, the strain rates in this type of simulation are typically much higher than those used experiments [36]. An additional set of simulations with a $10^8$ s$^{-1}$ engineering shear strain rate was performed for each sample to verify that the conclusions made in this work are not qualitatively affected by strain rate.

## 3. Results

### 3.1. Dislocation emission

Tensile deformation of the simulation cell induces dislocation emission from the grain boundary complexions. Figure 4 shows the moment of dislocation emission from the CGB, OGBC, and AIF samples. For all of the samples, Shockley partial dislocations that correspond to the {111}<112> slip system were observed, with some of these forming Lomer-Cottrell locks (stair-rod dislocations). The engineering tensile stress and strain were followed during the simulation and are presented in Figure 5(a). The critical stress was determined at the moment when the first dislocation was emitted. As the dislocation emission leads to decrease of the stress, the critical emission stress is equal to the maximum stress on the corresponding stress-strain curve. The average critical tensile stresses and resolved shear stresses are presented in Figure 5(b) for each system. The obtained shear stresses are close to, and even above for the OGBC and AIF samples, the ideal shear strength extracted by Ogata et al. for *ab initio* calculations in single crystal copper ($\tau_{ideal} = 2.16$ GPa) [37]. However, dislocation nucleation in an MD tensile test will be biased by (1) the presence of normal stresses and (2) the fact that interatomic potentials are approximations [38, 39]. A more reliable comparison can be made with a single crystal having an orientation of the inner grain (see Figure 2(a)) that is strained along the X axis under identical MD conditions. In this case, we measure that the critical resolved shear stress at the point of dislocation nucleation is 5.26 GPa. In Figure 5(b), the critical stress is lower for the OGBC and AIF samples, meaning that both Zr segregation to grain boundaries and amorphous complexion formation promote dislocation emission. In contrast, some other studies on the same topic have shown that solute addition to grain boundaries can suppress dislocation emission. For example, in the recent work of Borovikov et al. [19], which studied dislocation emission from a Σ11 grain boundary in the Cu-Ag system, a significant (up to two times) increase in the yield stress for both a Cu bicrystal with Ag solutes and an Ag bicrystal with Cu solutes was found. In another MD study focused on nanocrystalline Al-Pb alloys, Jang et al. [40] showed that dislocation emission is suppressed by Pb addition.

To rectify these contrasting reports of the effect of grain boundary doping on dislocation emission, we turn to the concept of excess grain boundary free volume. Previous studies have found an inverse relationship between grain boundary free volume and yield stress, demonstrating that solute addition to grain boundaries leads to a decrease in grain boundary free volume and an associated increase in yield strength for the alloys that were studied [41, 42]. Tucker and McDowell [18] observed this same inverse dependence in pure Cu samples with and without nonequilibrium grain boundary structures. These authors found that tensile strength decreased as the grain boundary state moved further from equilibrium, with dislocations being preferentially emitted from grain boundaries with a large excess free volume. It was concluded that the excess free volume promotes an atomic reordering that is necessary for the dislocation emission process.

To understand if this behavior can explain our results, we first measured the atomic volume per atom in the interfacial region during the tension simulations. Interfacial atoms were identified as atoms with unknown local crystal structure in the adaptive CNA analysis. The average atomic volume was obtained using Voronoi cells [43] and is presented in Figure 6(a). In the beginning of the test, at 0% tensile strain, the OGBC and AIF atomic volumes are almost equal, with both noticeably greater than the atomic volume of the CGB. During the tensile deformation, the AIF atomic volume increases more rapidly than the OGBC atomic volume, leading to a clear separation by ~3% tensile strain. Similar to the critical stress for dislocation emission, the critical atomic volume was measured at the moment when the first dislocation was emitted. The atomic volume of bulk copper (12.1 Å$^3$) was subtracted from these values to determine the amount of free volume in the interfacial region. Figure 6(b) plots the critical stress for dislocation nucleation versus the critical free volume at the interface, demonstrating an inverse dependence of the emission stress on free volume per atom. Consequently, we can conclude that OGBCs and AIFs promote dislocation emission because of an increase in excess of free volume that is caused by Zr segregation.

### 3.2. Dislocation propagation

With the effect of complexion structure on dislocation emission understood, we next moved to understand how subsequent propagation was affected. Figure 7 shows the CGB sample with a moving dislocation, with atoms colored according to the local crystal structure analysis. Shockley partial dislocations in Figure 7 appear as the lines bounding the stacking fault (red atoms). The CGBs appear as the vertical lines of gray atoms in this figure. The overlaid solid black lines indicate the limits of the inner grain, which has a width of ~17 nm. As every dislocation dissociated into leading and trailing partial dislocations, the time-dependence of the center-of-mass of the stacking fault was computed to allow for the characterization of dislocation propagation and to measure an average dislocation velocity. We use only the region of the stacking fault with a size of 4 nm in the middle of the grain, indicated by the dashed black lines in Figure 7, where the stacking fault remains relatively straight. The dislocation position at this time is denoted by a black dot. Each dislocation position was measured relative to its initial position, with a positive value indicating motion in the positive Y-direction and a negative value indicating motion in the negative Y-direction.

Dislocation positions as a function of time are shown in Figure 8, together with the corresponding engineering stress-strain curves. The solid lines in Figure 8(a-c) show the shear stress of the entire system, while the dashed lines represent the shear stress only in the inner grain where the dislocation is propagating. The rise in total stress, which is observed for all the samples, is caused by progressive elastic loading of the outer grain, which has no dislocations and therefore cannot plastically deform. This rise continues until new partial dislocations are emitted in the outer grain near the end of the test, near ~9% applied shear strain. In contrast, the shear stress in the inner grain slowly increases and reaches a plateau associated with steady-state dislocation propagation at a constant velocity.

Figure 9(a) shows stress-strain curves for the inner grains of one sample with CGBs and one sample with AIFs. The major peaks and valleys on the stress-strain curves in the samples are situated at approximately the same positions, corresponding to times when the two dislocations in the simulation cell interact through their stress fields. More important is the consistent shift upward of the shear stresses for the AIF sample. In Figure 9(a), the average shear stress during flow, measured by averaging the stress over the range of 3-9% shear strain corresponding to steady-state dislocation propagation, is denoted by a solid colored line. Average dislocation velocity was obtained by taking the linear slope of the dislocation position in the time range of 0.03-0.09 ns (i.e., the same range as the measurement for flow stress). Both the flow stress and dislocation velocity measurements were averaged over the six identical configurations run for each type of grain boundary sample, with standard deviations from the mean values also calculated. The extracted mean values and the corresponding standard deviations of the flow stress and dislocation velocity are shown in Figure 9. As one can see, the average flow stress is significantly higher for the samples with the Zr addition (OGBC and AIF), while the average dislocation velocity demonstrates a relatively constant value independent of complexion structure. That means that the OGBC and AIF samples require a 20-30% higher stress to sustain dislocation propagation with the same velocity as the CGB sample. The AIF sample is noticeably stronger than the OGBC sample, demonstrating that the amorphous film at the interface restricts dislocation propagation in some way.

Simulations with a lower shear strain rate of $10^8$ s$^{-1}$ support our findings as well. Figure 10 shows the simulations results for one configuration of each sample continuously deformed at a $10^8$ s$^{-1}$ strain rate. Figure 10(a) demonstrates the same strengthening effect that was observed previously in Figure 9(a), where the curves for the samples with OGBCs and AIFs are shifted upwards compared to the CGB sample. However, for this slower strain rate, we observe behavior that more clearly shows dislocation pinning and unpinning. The center section moves forward slowly as shear stress increases, until a critical value is reached and rapid propagation of the dislocation occurs as the dislocation edges unpin from grain boundary sites. The shear stress in the inner grain demonstrates a sharp decrease, then the process repeats itself. After ~8% applied shear strain, dislocations are emitted into the outer grain and we can no longer continue this analysis. The dislocation velocities are comparable for the three samples, consistent with findings for the $10^9$ s$^{-1}$ strain rate simulations, although the velocities are one order of magnitude lower with a value of ~80 m/s, as would be expected for a slower strain rate. The average peak stress for each sample is shown in Figure 10(c). This data follows the same trend as the data shown in Figure 9(b), proving that the general trends reported here are not sensitive to applied strain rate.

With an understanding of the effect on the critical flow stress for propagation, we return to find a mechanistic explanation behind these trends. Returning to the raw simulation data, the wavy nature of the shear stress-strain curves (see dashed lines in Figures 8(a-c),9a and 10a) indicate that dislocation propagation might be impacted by a dislocation pinning/unpinning process at the grain boundaries. This process is followed by an abrupt motion forward, which appears smooth in Figure 8(d-f) (except for the complete stop) because of the arched shape of the partial dislocations due to pinning at the ends. The bowing of the partial dislocations gives the impression that the entire dislocation moves in a very smooth manner, when in fact the parts of the dislocation near the boundaries at the edge of the inner grain move through discrete jumps due to pinning and unpinning.

Figure 11(a) shows the pinning of the left side of the partial dislocations along the grain boundary, while the middle and right side continue to move. This pinning leads to the bowing of both the leading and trailing partial dislocations. Dashed lines mark the location of the pinning sites, while inspection of the various time steps in Figure 11(b) shows that process in more detail. The atomistic structure of these pinning sites shows that they correspond to grain boundary ledges, which are marked by solid red and blue lines in Figure 11(b). At 7 ns, the leading dislocation was stopped by a ledge (red circle), while the trailing dislocation continued to move, leading to a reduction in the width of the stacking fault. At 9 ns, the trailing dislocation reaches a ledge (blue circle) and both of the dislocations remain completely stationary until ~13 ns, when the leading dislocation became unpinned again. This process continues as the simulation progresses. A similar mechanism of dislocation pinning by ledges in local grain boundary structure was described by Van Swygenhoven et al. [11]. In this study, we find that ledges are significantly larger on average in the AIF sample, going deeper inside the inner grain because of the disordered nature of this grain boundary complexion. Such large ledges are expected to create a higher energy barrier for the dislocation propagation near the interface. In this case, one may use the geometric roughness as a proxy for the roughness of the energy landscape. The average measured boundary roughness parameters were 6% for CGBs, 8% for OGBCs and 16% for AIFs. Boundary roughness is increased more than two times in the AIF sample, which can make propagation much harder and lead to the increase in the required critical applied stress. On the other hand, the strengthening effect of the OGBCs cannot be readily attributed to the ledge size, as CGBs and OGBCs have relatively similar grain boundary structures.

To understand how the Zr dopants affect dislocation propagation, we also investigated the local atomic stress distribution around the moving dislocations. The local atomic structure of the slip planes is shown in Figure 12 together with the atomic shear stress distributions, resolved along the ZX direction which would drive dislocation movement. As compared to the CGB sample, the samples with OGBCs and AIFs demonstrate an increase in the dislocation curvature, due to the increased bowing of the dislocation. In addition, the atomic shear stress distribution near the interfaces in the OGBC and AIF samples have multiple areas of reduced stress, with a few examples indicated by the black arrows in Figure 12(e) and (f). While such reduced stress areas are present in all three samples, but the OGBC and AIF samples contain larger regions that are more densely distributed when compared to the CGB sample (Figure 12(d-f)). Dislocation propagation is driven by the stress gradient around the partial dislocations, with positive stress in front and negative stress in the rear. The areas of the reduced stress near the grain boundary complexion interfaces therefore act as pinning areas by locally reducing this stress gradient.

Since the regions of reduced stress areas are absent in the CGB sample, one can conclude that they are caused by Zr segregation. First, solute segregation to grain boundaries can lead to local changes in the elastic properties of the interfacial material. In addition, the atomic radius mismatch between the alloying element and matrix atoms may locally modify the local stress field. Since local stresses at the boundaries can either magnify or reduce the dislocation's stress field, these local changes will affect propagation.

## 4. Discussion

By comparing our observations with experimental reports, we can gain insight into the critical mechanism that limits plasticity in nanostructured alloys. To facilitate such a comparison, the relative changes of the critical emission stress and of the flow propagation stress were calculated, using the values for the CGB sample as a baseline. This measurement was chosen to allow for a fair comparison between the values taken from each mechanism, but also for a comparison with experimental reports. Experimental values were taken from Khalajhedayati et al. [5], with the pure nanocrystalline Cu sample also serving as the baseline for this dataset. The relative changes are presented in Figure 13, where a negative relative change denotes softening and a positive relative change denotes strengthening. Dislocation emission is easier with the addition of either OGBCs or AIFs, decreasing the critical stress required for activation. In contrast, the MD simulations of dislocation propagation demonstrate the strengthening effect of OGBCs and AIFs. The experimental microcompression data from the nanocrystalline Cu and Cu-Zr alloy indicates strengthening caused by both solute segregation to ordered grain boundaries as well as the addition of AIFs. The experimentally measured yield strengths were 0.74 GPa for the CGB sample, 0.938 GPa for the OGBC sample, and 1.086 GPa for the AIF sample. It is worth noting that the experimental data points do not all have the exact same grain size. The listed under OGBC and AIF had grain sizes of 45 nm, which was slightly larger than the as-milled grain size of 30 nm for the sample used as the baseline. Correcting this trend for grain size would further accentuate the strengthening effect from the addition of OGBCs or AIFs. However, the best method for making this correction is unclear and we are mainly interested in the general trend, so we do not attempt to alter the data here.

The qualitative comparison of our MD results with the experimental data suggests that dislocation pinning by grain boundary complexions during dislocation propagation is the rate-limiting deformation mechanism in nanocrystalline Cu-Zr alloys. The bicrystal geometry considered in this study does not clarify how dislocation nucleation and propagation are affected by more specific features such as grain boundary character, curvature, faceting, etc., nor does it treat the importance of triple junctions as local stress concentrations. These features, as well as possible concurrent interface-mediated processes such as grain boundary sliding, may alter the critical stresses required for dislocation nucleation and propagation in nanocrystalline metals and alloys. However, modeling a polycrystalline sample would make it impossible to isolate the two mechanisms one-by-one. Furthermore, the general conclusions we have drawn (e.g., inverse relation between the critical nucleation stress and the interface free volume) should apply to a discussion of other interfacial sites as well, including other types of grain boundaries and even triple junctions. The results shown here do demonstrate a clear connection between complexion type and the mechanisms which control plasticity of nanocrystalline metals with grain sizes of several tens of nanometers.

Moreover, experimental and MD studies in the literature which focus on the influence of the grain boundary solute segregation on the nanocrystalline material strength for other binary systems provide support for this conclusion as well. For example, Rupert et al. [20] developed a phenomenological model that is based on the idea of dislocation pinning at nanocrystalline grain boundaries and successfully applied it to explain solid solution effects on strength in nanocrystalline Ni-W, Pt-Ru, Ni-Co, Ni-Cu, and Ni-Fe alloys. While this work was focused on how solutes alter the properties of the grain interior, it was based on dislocation propagation as the critical event, similar to our conclusion here. In addition, Schäfer et al. [44] found that the yield stress for different compositions in a Pd-Au alloy did not correlate with the resistance to dislocation emission, meaning that the dislocation emission cannot be rate-limiting in the nanocrystalline Pd-Au alloy either.

Additional evidence for a mechanism of dislocation pinning to grain boundary complexions by solute atoms can be found in literature. For example, Vo et al. [42] investigated the strengthening of nanocrystalline Cu by grain boundary doping with Nb, Fe, and Ag atoms. Vo et al. found that the strength of nanocrystalline Cu-based alloys increased proportionally to the atomic size mismatch between Cu and the dopants. According to our hypothesis, solute atoms with a relatively large size would create a strong local stress field and have a stronger influence on the stress gradient, making the pinning areas at grain boundary complexions more attractive. An unexpected consequence of this mechanism is that grain boundary-dislocation pinning would only depend on the solute distribution in the layer very close to the border between the grain and the intergranular film. Consequently, the dislocation pinning process would not be expected to be affected by AIF thickness, although this hypothesis remains untested.

## 5. Conclusions

In this work, the influence of ordered and disordered grain boundary complexions on dislocation emission and propagation was determined using MD and hybrid MD/MC simulation methods. Several important conclusions were made and are presented below.

- Dislocation emission was easier in the OGBC and AIF samples than in the CGB specimen. The critical stress for dislocation emission demonstrates an inverse correlation with the excess free volume at the interface, meaning the increase in the free volume due to Zr segregation to the grain boundary complexions is responsible for the softening effect observed in our simulations.

- In contrast, the flow stress for dislocation propagation is larger for the OGBC and AIFs samples than for the CGB sample. This strengthening effect is caused by a combination of local ledges in boundary structure and local stress variation cause by Zr dopant segregation.

- Comparison of our MD results with available experimental data suggests that, while both mechanisms can be important for nanocrystalline plasticity, dislocation propagation past grain boundary pinning sites is rate-limiting. Complexion addition strengthens nanocrystalline materials, in line with our observations for dislocation propagation.

Taken as a whole, our work suggests that the strength of nanocrystalline metals with grain sizes in the tens of nanometers is controlled by grain boundary-dislocation pinning. Dopant segregation and grain boundary complexion formation can make the dislocation pinning effect stronger, creating areas of reduced stress near the grain boundary interfaces. The mechanisms studied in this work reveal important atomic details of nanocrystalline plasticity and allows materials engineers to predict the effect of the different alloying elements on a nanocrystalline material's strength.

**Acknowledgements:**

This research was supported by the U.S. Army Research Office under Grant W911NF-16-1-0369.

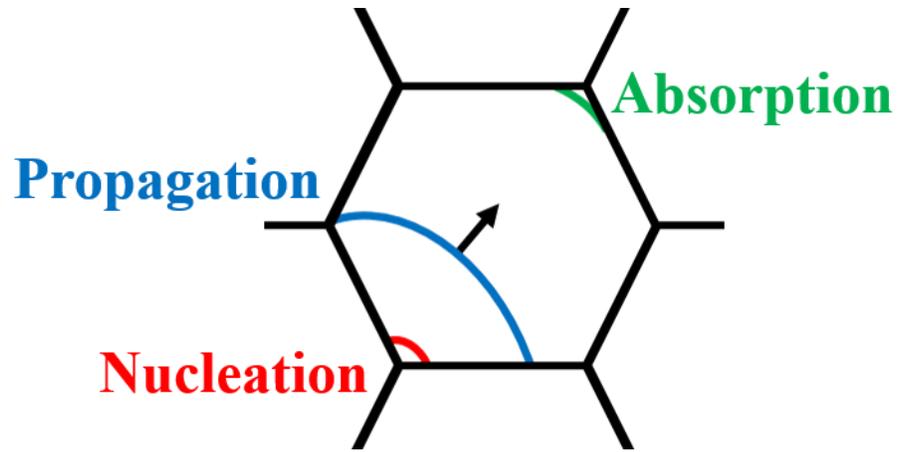

Figure 1. Schematic representation of the three important grain boundary–dislocation interactions during nanocrystalline plasticity: nucleation from (red curve), propagation along (blue curve) and absorption at (green curve) the grain boundary (black hexagon).

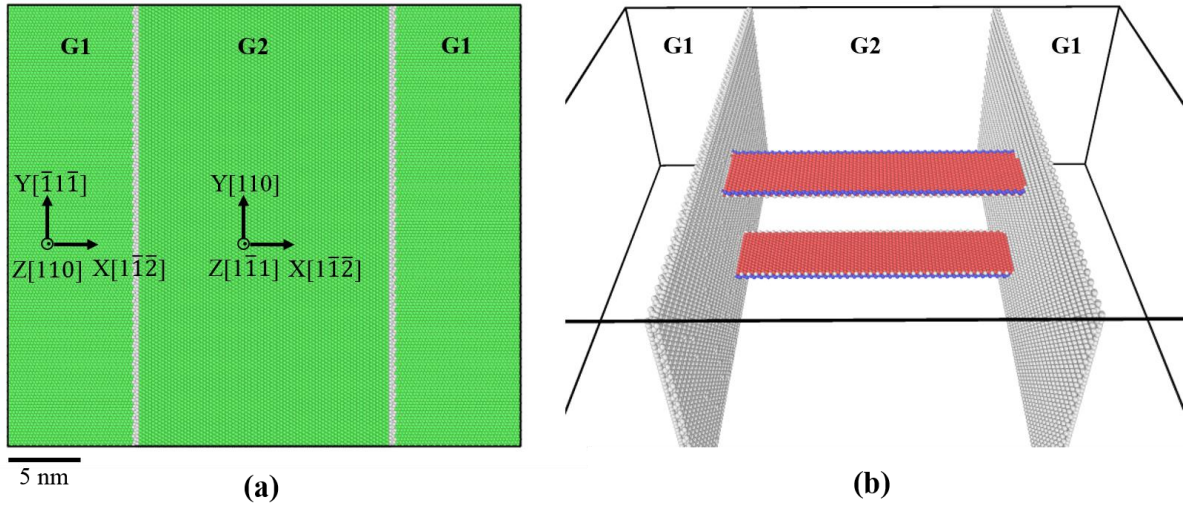

Figure 2. (a) A schematic representation of the initial geometry of the samples used to study dislocation emission and dislocation propagation. (b) The simulation box for the dislocation propagation study, with only non-fcc atoms shown. Atoms are colored by their local crystal structure: green – fcc, blue – bcc, red – hcp, and gray – unknown.

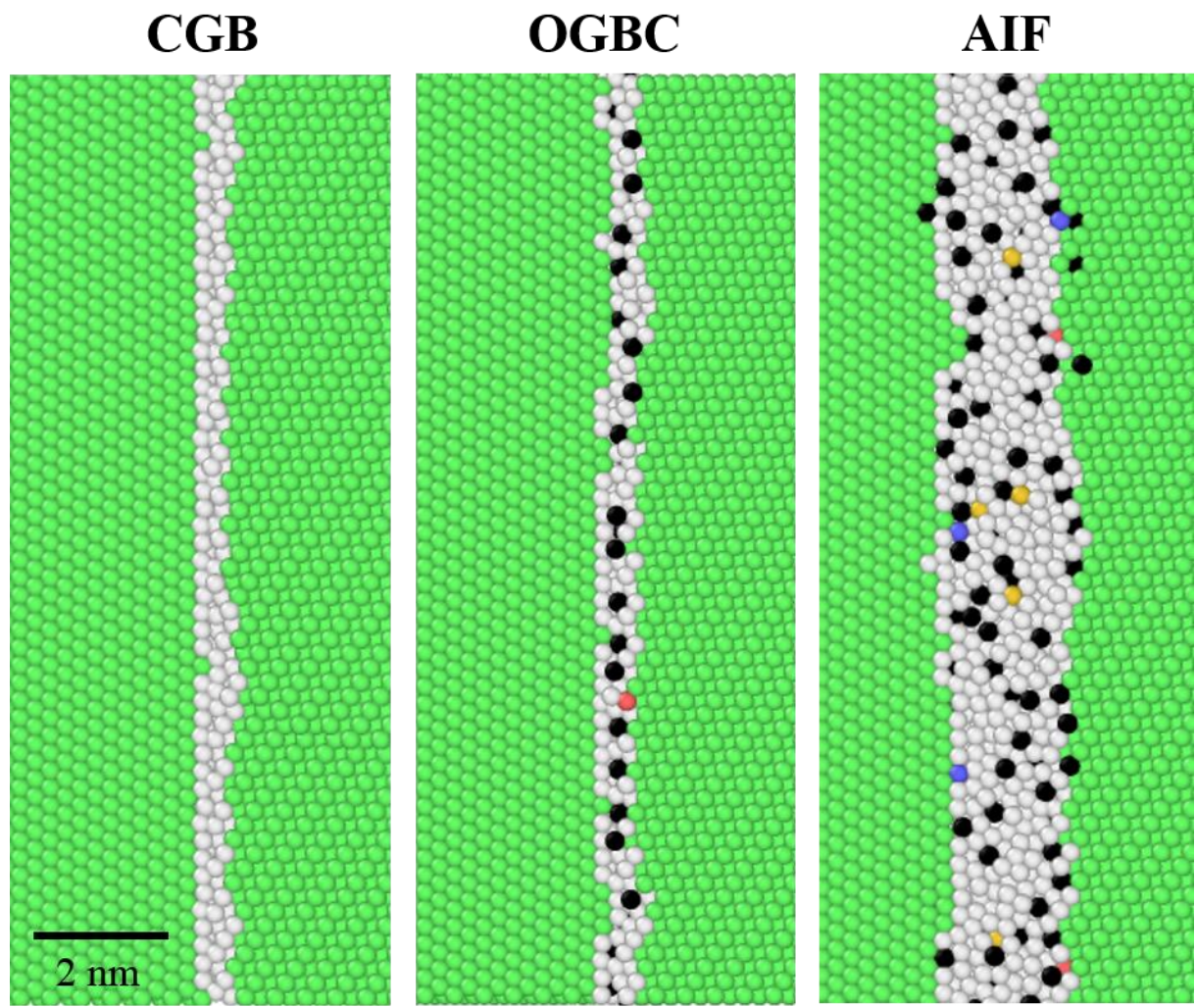

Figure 3. Equilibrium grain boundary structures obtained using the hybrid MD/MC method: clean grain boundary (CGB), ordered grain boundary complexion (OGBC) and amorphous intergranular film (AIF). Atoms are colored in relation to their local atomic structure, while Zr atoms are shown as black spheres.

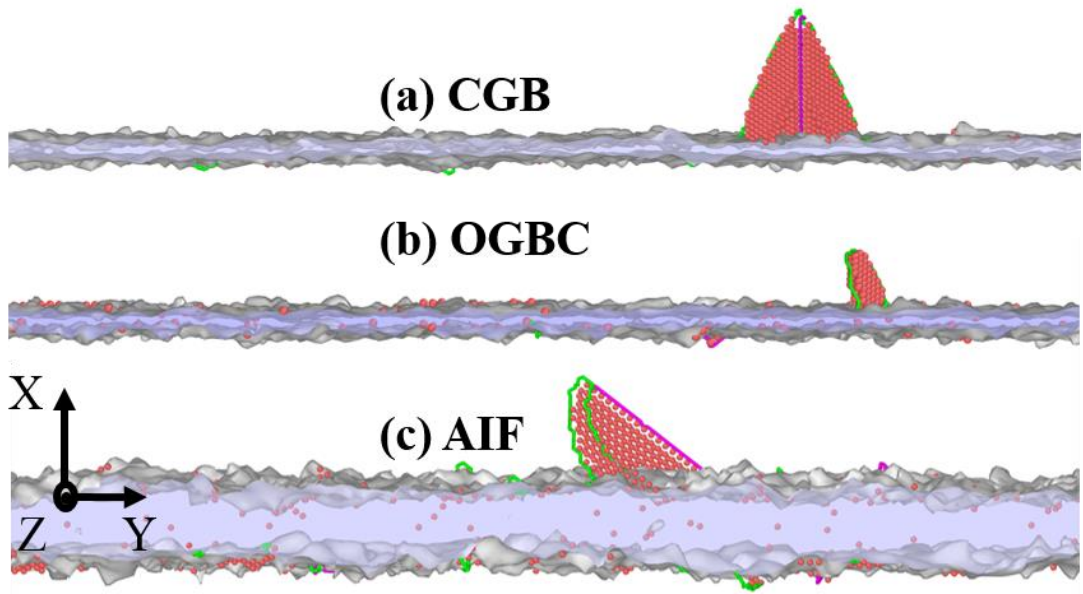

Figure 4. Dislocation nucleation from different types of grain boundary complexions, analyzed by the DXA method. The gray background represents the defect mesh and the emitted dislocations are colored by type. Shockley partials are shown as green, while stair-rod dislocations appear as magenta. Red spheres represent atoms in stacking faults.

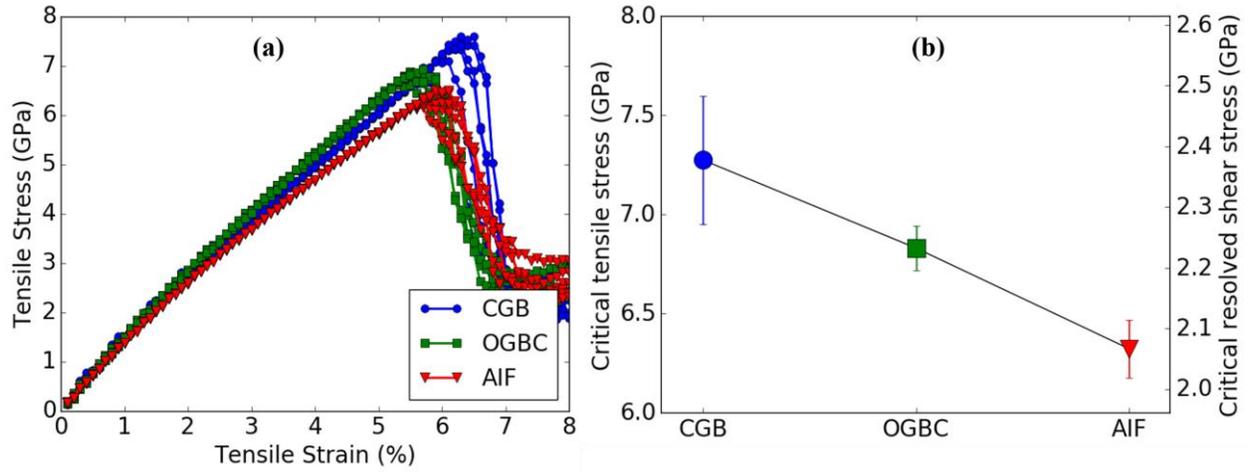

Figure 5. Data from dislocation emission simulations. (a) Tensile stress-strain curves for the different grain boundary complexions, with multiple initial configurations shown. (b) Critical stress values taken when the first dislocation is emitted from the each grain boundary complexion.

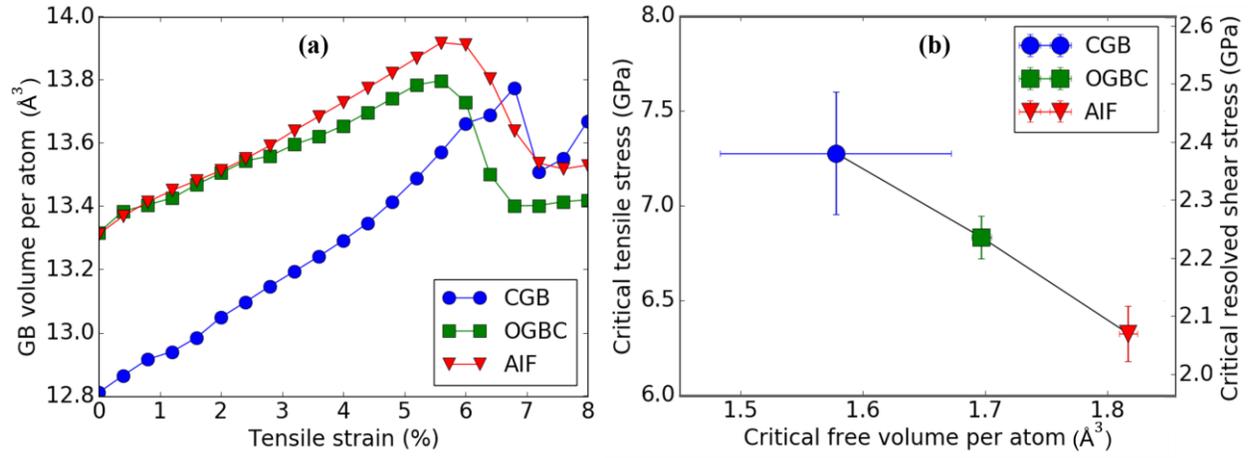

Figure 6. (a) Evolution of the average grain boundary (GB) volume per atom during tension in the dislocation emission simulation. (b) The critical emission stress plotted against the critical grain boundary free volume per atom for all studied grain boundary complexions.

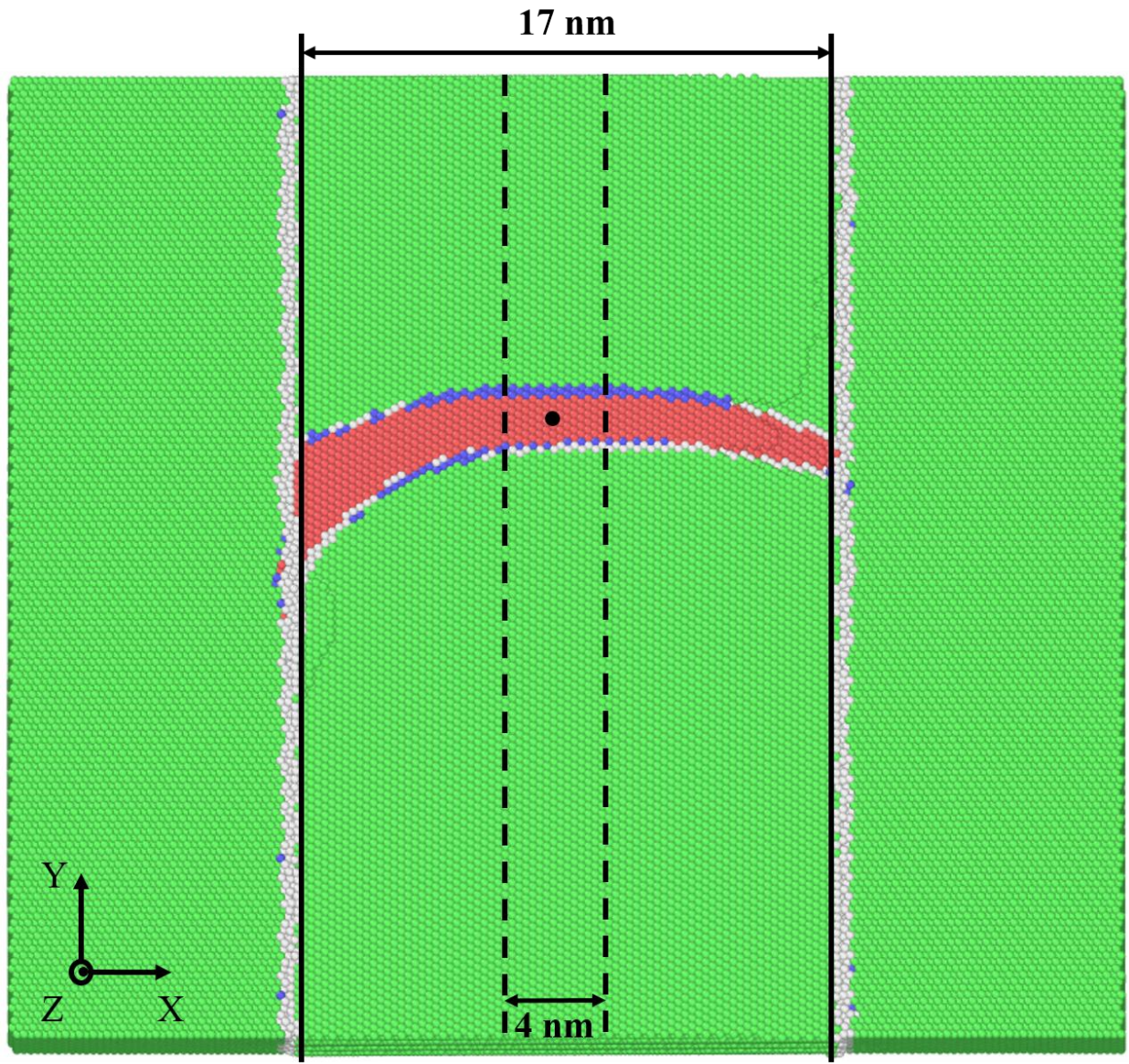

Figure 7. Atomic snapshot along the slip plane during the dislocation propagation simulation, where atoms are colored by their local atomic structure. The black solid lines show the limits of the inner grain and the dashed lines indicate the region where the dislocation position is measured (location in this snapshot is marked by the black dot).

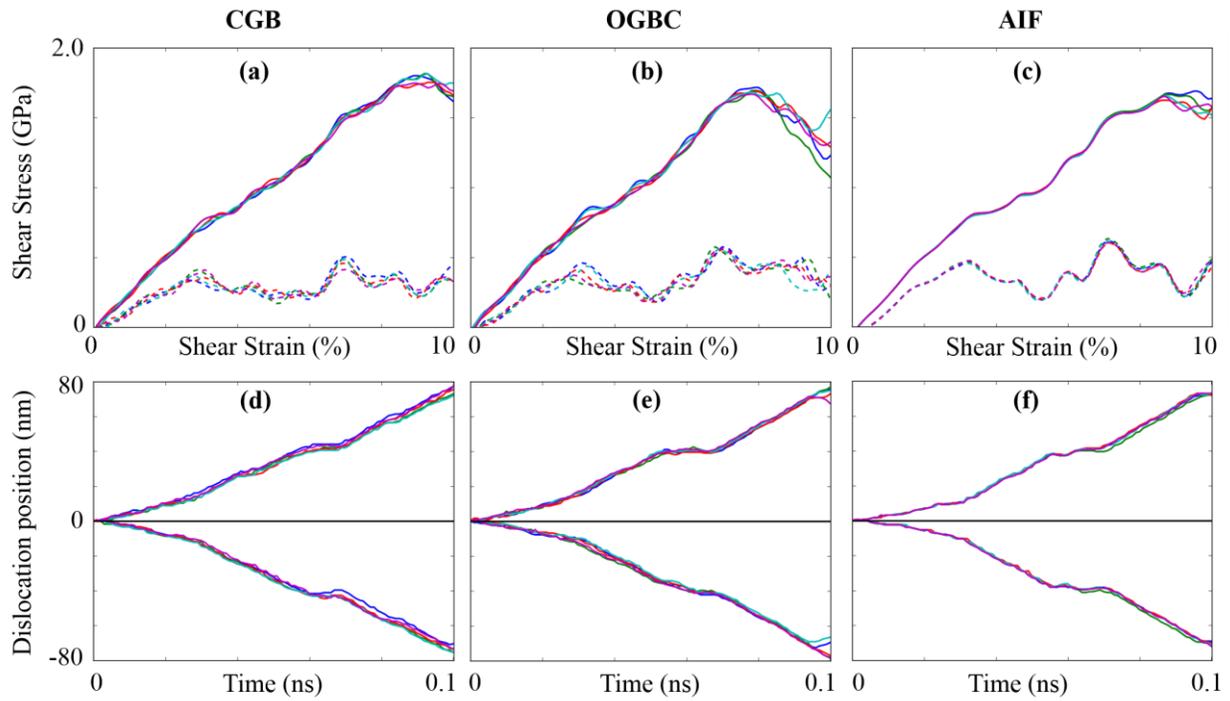

Figure 8. (a-c) Shear stress-strain curves and (d-f) dislocation positions as a function of time for the samples with the different grain boundary complexions during dislocation propagation. Different line colors represent different initial configurations.

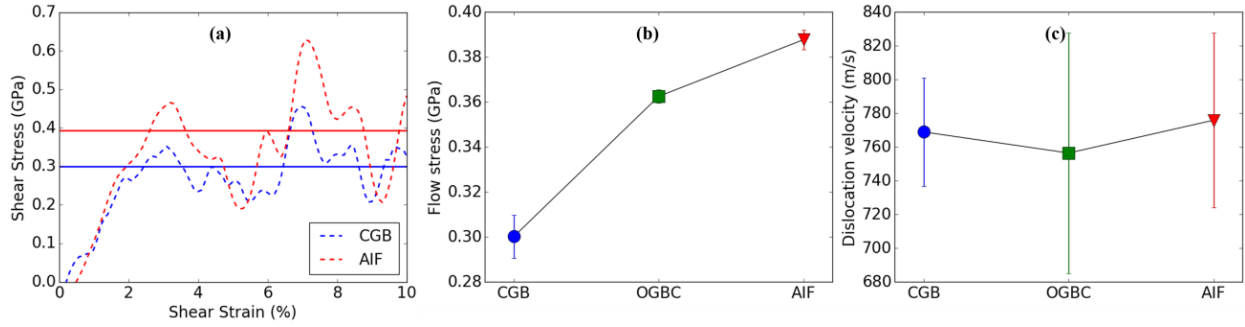

Figure 9. (a) Shear stress-strain curves for the inner grains in the samples with the different grain boundary complexions that were deformed with a $10^9$ s$^{-1}$ engineering shear strain rate. (b) Flow stress for dislocation propagation and (c) average dislocation velocity for samples with the different grain boundary complexion types.

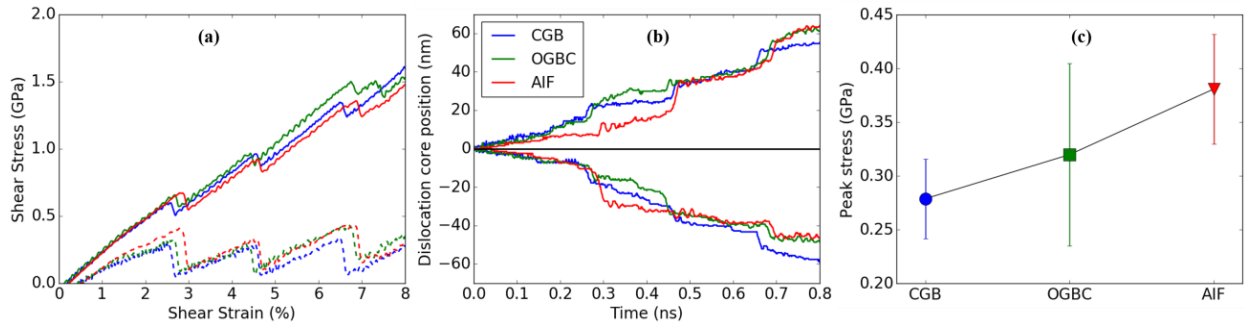

Figure 10. (a) Shear stress-strain curves, (b) dislocation positions as a function of time and (c) average peak stresses for the samples with the different grain boundary complexions. The samples were deformed with a slower engineering shear strain rate of $10^8$ s$^{-1}$. Different line colors represent different samples: blue – CGB, green – OGBC and red – AIF.

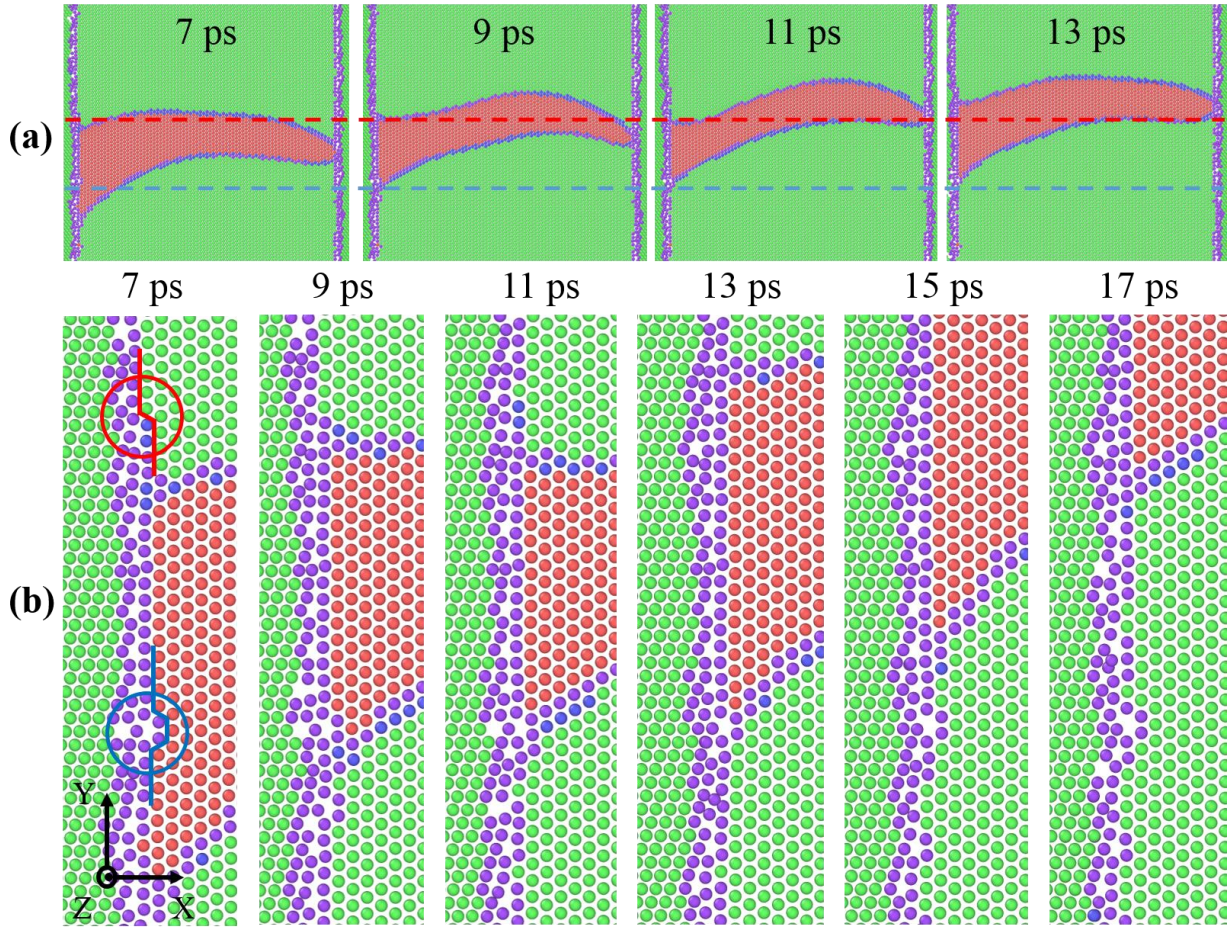

Figure 11. (a) Pinning, bowing, and unpinning of the partial dislocations in the CGB sample. Dashed lines represent the positions of the pinning sites. (b) Atomic snapshots demonstrate dislocation pinning by ledges in the local grain boundary structure along a CGB. The CGB may be identified as a vertical column of the violet atoms with an unknown local atomic structure (violet was chosen instead of the gray color used in other figures for better visibility). The partial dislocations move up and may be identified as the diagonal lines between the stacking fault (red spheres) and crystalline Cu (green spheres). The red circle represents a ledge that is pinning the leading dislocation, while the blue circle represents a ledge that is pinning the trailing dislocation. The red and blue solid lines show the shape of the grain boundary ledges.

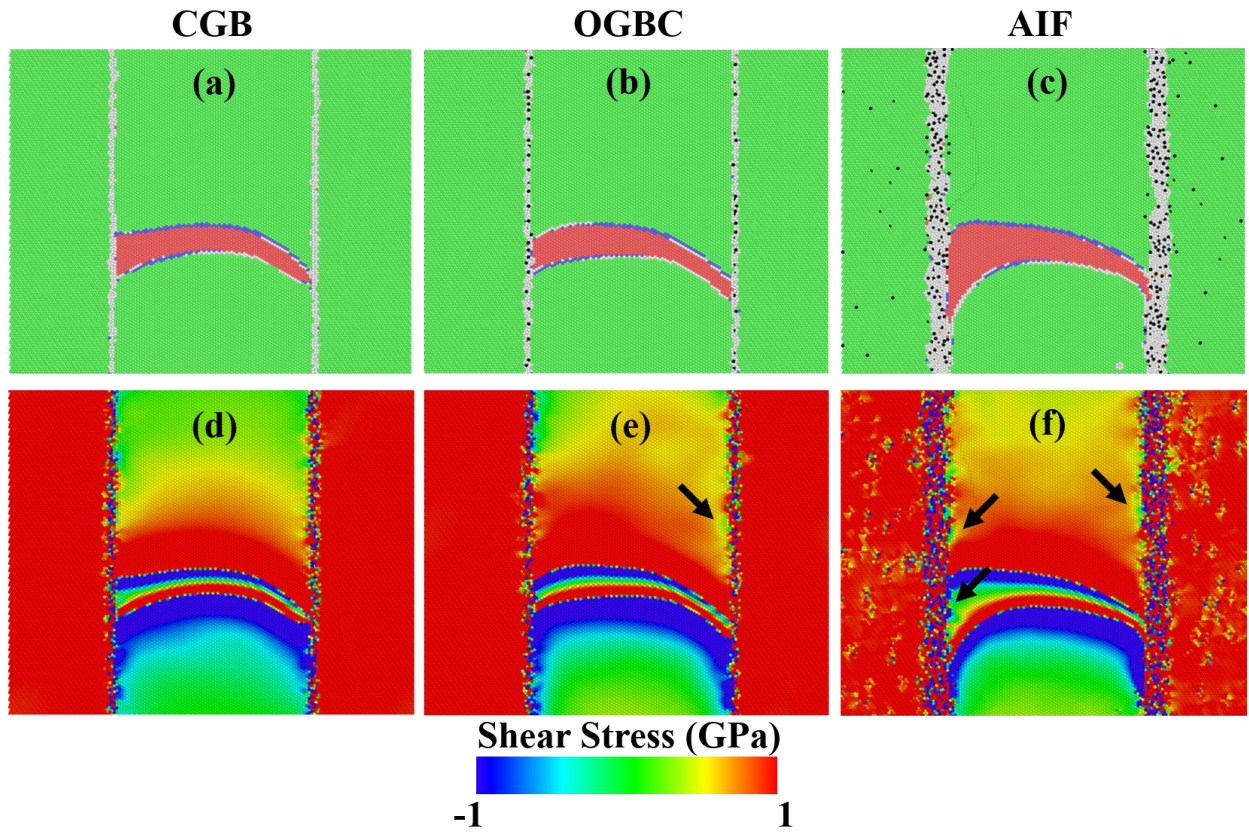

Figure 12. (a-c) Local atomic structure and (d-f) atomic shear stress distribution during dislocation propagation simulations at 3% applied shear strain. Black arrows denote local regions of low stress.

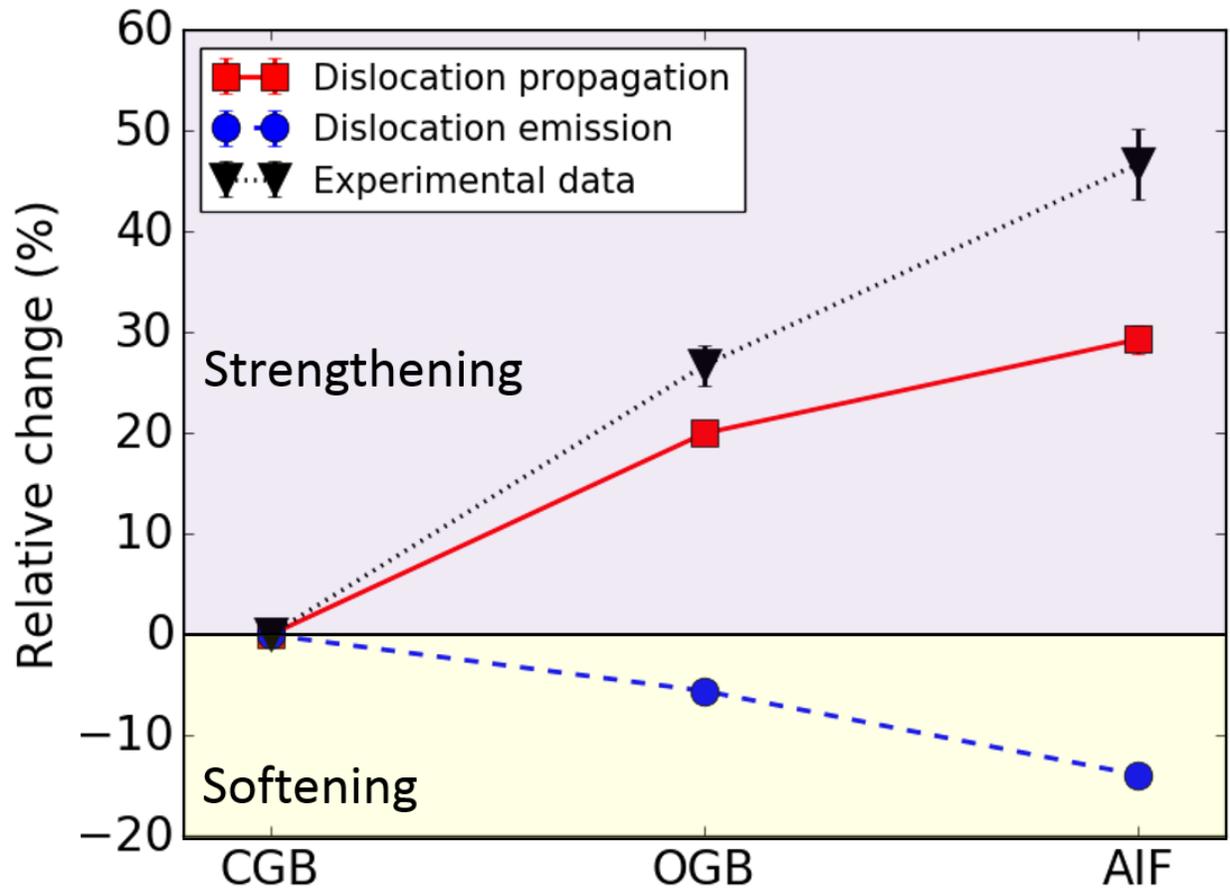

Figure 13. Relative changes of the critical stress required for dislocation emission and propagation, as measured by the MD simulation in this work. The relative change in the yield stress obtained from the experimental microcompression data for nanocrystalline Cu-Zr in Ref. [5] is also shown for comparison.